\documentstyle{article}
\newcommand{\q}{\quad}
\def\\q.{\quad .}
\newcommand{\beq}{\begin{equation}}
\newcommand{\eeq}{\end{equation}}

\newcommand{\bdm}{\begin{displaymath}}
\newcommand{\edm}{\end{displaymath}}
\newcommand{\re}[1]{(\ref{#1})}

\newcommand{\goesto}{{\scriptstyle \rightarrow}}

\begin{document}
\title{Radiation from a Charge Uniformly Accelerated for All Time} 
\author{Stephen Parrott \\
\\
Department of Mathematics and Computer Science\\
University of Massachusetts at Boston\\ 
100 Morrissey Blvd.\\
Boston, MA 02125
\\
USA
\\ 
}
\maketitle
\begin{abstract}
A recent paper of Singal \cite{singal} argues that a uniformly
accelerated particle does not radiate, in contradiction to the 
consensus of the research literature over the past 30 years.
This note points out some questionable aspects of Singal's argument
and shows how similar calculations can lead to the opposite conclusion.  
\end{abstract}
\section{Introduction} 
\newcommand{\calE}{{\cal E}}
\newcommand{\bfE}{{\bf E}}
\newcommand{\bfB}{{\bf B}}
\newcommand{\bfn}{{\bf n}}
\newcommand{\bfbeta}{{\bf \beta}}

Over 40 years ago, it was a matter of controversy  
whether radiation would be observed from a charge which
had been uniformly accelerated for all time.  
The situation around 1960 is described in \cite{fulton/rohrlich},
which concludes that ``contrary to claims in some standard sources,
(Pauli, Von Laue), a charge in uniform acceleration does radiate''. 
Subsequently a consensus for radiation seemed to have been reached
(I know of no paper claiming the contrary within the past 30 years)
until a recent paper of Singal 
presented a calculation which it interprets as proving that 
``there is no electromagnetic radiation
from a uniformly accelerated particle'' 
\cite{singal}[p. 963].

This note analyzes this calculation and Singal's interpretation. 
It observes that Singal's method applied to a particle uniformly
accelerated for a {\em finite}, but arbitrarily long, time leads
to the opposite conclusion: such a particle {\em does} radiate
in accordance with the Larmor law.  The answer given by Singal's method
to the radiation question for uniform acceleration for {\em all} time
is not a limit of the corresponding answers for acceleration for
arbitrarily large times.    
This inconsistency could be viewed as a paradox or as an indication
of the unreliablity of Singal's method.   

Other inconsistencies of the same nature have been 
long known. 
All such inconsistencies known to this author 
can be traced to mathematical 
ambiguities in formulating the radiation problem for a particle
uniformly accelerated for {\em all} time, as opposed to
the mathematically unambigous problem of determining whether
radiation would be observed from 
a particle uniformly accelerated for a very long, but finite, time.  

We conclude by reviewing a related ``paradox'' 
(that calculation of radiated
energy-momentum 
by integration over Bhabha and Dirac tubes give the same answer for 
asymptotically free particles but different  answers for a particle
uniformly accelerated for all time).  The only known resolution to this
paradox is to disallow (for good reasons) these calculational methods for 
particles which are not asymptotically free.  We suggest that  
Singal's paradox should be similarly resolved.

\section{Summary of Singal's method and results} 
 
Singal's calculations are performed entirely within a fixed
Lorentz frame which we'll call the ``laboratory frame''.  
Coordinates in the laboratory frame will be denoted
$(t, x, y, z)$.

Consider a particle of charge $e$ whose worldline 
is described in laboratory coordinates as a function
of proper time $\tau$ by
\begin{equation}
\label{ap4.1}
\tau \mapsto g^{-1}(\sinh g \tau, \cosh g \tau, 0,0)
\end{equation}
This describes a particle moving on the $x$-axis with 
uniform proper acceleration $g$ in the positive $x$-direction,
so that the particle comes to rest
at $x = g^{-1} $ at time $t = 0$.
Singal calculates the energy $\calE$ in its electromagnetic field at
the laboratory time $t = 0$ using the usual expression 
$(\bfE^2 + \bfB^2)/8\pi$ for the three-dimensional field energy density: 
$$
\calE = \int_W d^3v \, (\bfE^2 + \bfB^2)/8\pi   
$$ 
where $\bfE$ and $\bfB$ are respectively the electric and magnetic fields.

Of course, the integral is expected to diverge unless 
the region of integration $W$ omits some ball
containing the particle, since the corresponding integral for
a stationary particle is $e^2/2r_0$ when a ball of radius $r_0$
centered at the particle is omitted. The ball which the paper chooses
to omit is a ball of some given positive radius $r_0 $  with the particle
at the center at laboratory time $t = -r_0$.   In Singal's terminology, 
the ball consists of all points at ``retarded distance'' 
no more than $r_0$ from the particle.   The ``retarded distance'' 
of $(t, x, y, z)$ from the particle is defined as the laboratory-frame distance   
to the point where a past-directed light ray starting at $(t,x,y,z)$
will intersect the particle's worldline.  
The region of integration $W$ is taken to be 
the set of all points whose retarded distance $R$ from the particle
satisfies $r_0 \leq R < \infty$.  

This integration region $W$ may
also be described as the open half-space $t = 0, x > 0$ with
the above ball of radius $r_0$ omitted.  Only a half-space 
is covered because points with $t = 0, x \leq 0$ cannot be 
connected to the wordline with a light ray.  
Despite this, the fields determined by the distributional
Maxwell equations as given in \cite{boulware}%
\footnote{See also \cite{bondi/gold} and earlier references
cited there and in \cite{fulton/rohrlich}.  We use \cite{boulware}
as a convenient modern reference for these fields.} 
do not vanish on $x = 0$ and Singal's decision
not to include them in the integration computing the field energy 
is one controversial aspect of his method.  
This will be discussed more fully below.

The result of Singal's calculation \cite{singal} 
is that the total field energy $\cal E$ as defined above is given by
$$
\calE = e^2/2r_0 \q ,
$$
which is the same as for a stationary particle. 
From this he concludes that no energy has been radiated up
to time $ t = 0$.

One might try to resolve the paradox by noting that 
there {\em is} infinite energy 
in this field if the ball of radius $r_0$ is not excised.
However, this resolution is intuitively unsatisfactory  
because according to the classical picture of electromagnetic energy
propagating at the speed of light (unity), 
any field energy within this ball must have been emitted  
between $t = -r_0$ and $t =0$.
If there is infinite energy in the fields
at $t = 0$ due to the infinite energy radiation from the infinite past 
to time 0, then at time 0 
there ought to be infinite energy in the electromagnetic
field {\em outside the ball of radius $r_0$}. That is, $\calE$ as 
calculated by Singal would still be expected to be infinite.  

\section{Singal's method applied to a particle uniformly accelerated 
for a finite time}

We shall suggest below that a particle uniformly accelerated for
{\em all} time represents a situation too mathematically singular
to be reliably treated with the sort of mathematical manipulations 
customary in this field.  
To clarify the singular mathematics associated with a particle
uniformly accelerated for all time, 
we apply Singal's method to  
a particle uniformly accelerated for only a finite time
which can be arbitrarily long. 
We shall see that standard, nonsingular mathematics leads
to the conclusion that the energy in the field at time $t = 0$
goes to infinity as the beginning of acceleration is pushed
back to the infinite past. 
In other words, according to Singal's criterion for radiation
(excess energy in the field at time 0 relative to the energy
of a Coulomb field),
a particle uniformly accelerated for a finite time {\em does} radiate. 

Now we begin the calculation of the energy in the field at $t = 0$
of a particle uniformly accelerated for a finite time
using Singal's method \cite{singal}, 
with which we assume the reader is familiar. 
Suppose the acceleration starts at laboratory time $ t = -R_0 < 0 $ 
in the distant past, 
$R_0$ being the retarded distance of the starting event,
and continues to time $t = R_0$ in the future, 
the particle being 
otherwise at uniform velocity.  The worldline during the interval
of uniform acceleration will be taken as  \re{ap4.1}, with
uniform velocity otherwise.  

Singal's equation (5) 
transforms the expression for the electric field
$\bfE$ given in \cite{jackson} to a sum of two orthogonal terms: 
\begin{eqnarray*} 
\bfE &=& \bfE_{radial} + \bfE_{trans} \\ 
&=& e \frac{\bfn}{\gamma^2 R^2(1 - \bfbeta \cdot \bfn )^2} 
+ e \frac{\bfn \times \{ \bfn \times (\gamma\bfbeta + \gamma^3 
\dot{\bfbeta} R)\}}{\gamma^3 R^2 ( 1 - \bfbeta \cdot \bfn)^3} 
\end{eqnarray*}
The notation is that of \cite{singal} and \cite{jackson} 
except that 
the velocity of light is taken as unity:  $\bfbeta$ is the  
particle's spatial vector velocity at the ``retarded'' point connected 
by a forward-pointing light ray to the field point at which
$\bfE$ is evaluated, 
$\gamma := (1 - \bfbeta^2)^{-1/2}$,
$\bfn$ is the spatial unit vector  pointing
from the space coordinates of the retarded point to the space coordinates
of the field point, 
and $R$ is the ``retarded distance'' defined as
the laboratory-frame distance of the retarded point from the field point. 
The ``radial'' first term of the right side will be abbreviated 
$\bfE_{radial}$, and the ``transverse'' second term $\bfE_{trans}$.
The magnetic field $\bfB$ (which is zero in the context of
\cite{singal} but not here ) is given by: $\bfB = \bf n \times E$. 

The field energy ${\cal E}$ is a sum 
\begin{eqnarray*}
{\cal E} &=& \int \frac{{\bf E}_{radial}^2}{8\pi}\, dv +  
\int \frac{{\bf E}_{trans}^2}{8\pi}\, dv  
+ \int \frac{{\bf B}^2}{8\pi}\, dv \\
&=& \int \frac{{\bf E}_{radial}^2}{8\pi}\, dv +  
2 \int \frac{{\bf E}_{trans}^2}{8\pi}\, dv  
\end{eqnarray*} 
The transverse term is easily integrated using Singal's volume element 
\cite{singal}[p. 964]
$$dv = 2\pi R^2(1 - \beta \cos \theta)\sin \theta \, dR \, d\theta \q ,$$ 
the result being:
\begin{eqnarray}
\label{etrans}
\int \frac{{\bf E}_{trans}^2}{8\pi}\, dv  &=& 
\frac{e^2}{8\pi} \frac{\beta_0^2}{\gamma_0^4}
\int_{R_0}^\infty dR \int_0^\pi d\theta \,  
\frac{ 2\pi R^2 (1 - \beta_0 \cos \theta) \sin^3 \theta}
{R^4(1 - \beta_0 \cos \theta)^6} 
\nonumber \\ 
&=& \frac{e^2g^2R_0}{3} 
\end{eqnarray} 
where $\beta_0$ and $\gamma_0$ are respectively the velocity and
corresponding $\gamma$-factor for the motion before the acceleration
started, and 
we have used the relation $\gamma_0^2 = 1 + g^2 R_0^2$
to express the result in terms of $R_0$. 
The contribution of the magnetic field to the field energy
is the same as \re{etrans}. 
The integral of the square of the radial term is 
(due to a fortuituous cancellation of $\gamma $ factors in
the integrand) identical, both in calculation and result,
with Singal's calculation for a particle uniformly
accelerated for all time.
The final result is:
\begin{eqnarray}
\calE &=& 
\int \frac{{\bfE^2} + \bfB^2}{8\pi}\, dv \\
&=& \frac{e^2}{2r_0} + 
\frac{2e^2 g^2 R_0}{3} 
\end{eqnarray}

The important point is that $\calE \goesto \infty $ as $R_0 \goesto \infty$.
If we take the result for a particle uniformly accelerated forever as
a limit of this result as $R_0 \scriptstyle \rightarrow \infty$, 
then the field energy in all space at time $t = 0$, 
excluding the above
ball of radius $r_0$ surrounding the particle, 
is infinite as expected.  

\section{Discussion of Singal's paradox} 

We propose that a reliable calculational method should 
have {\em at least} the following property:  the answer to the question 
\begin{quote}
``Does a particle uniformly accelerated for all time radiate?'' 
\end{quote}
should be the same as the answer to 
\begin{quote}
``Does a particle uniformly accelerated for a finite but arbitrarily
long time radiate?'' 
\end{quote}
The preceding section demonstrated 
that Singal's method does not have this property.

Any method without this property
is of questionable applicability to observational physics.  
Any physical meaning for uniform acceleration for all time 
must in practice be derived 
from approximations by acceleration for finite times.
We will never observe a particle uniformly accelerated for {\em all}
time, but we can hope to observe particles uniformly accelerated for
very long times.

The above argument that Singal's method is unreliable
does not tell us {\em why} it is unreliable.  What specific
feature of these plausible calculations is suspect?

The most likely culprit seems Singal's decision to limit the region $W$ 
of integration to the half-space $ t = 0, x > 0$.
Superficially this seems reasonable, since these are the only points 
on the hyperspace $t = 0$ which are causally
connected to the particle---no light ray from the particle can reach  
the two-dimensional plane $ t=0, x = 0$  (and also no light ray from
such a point can reach the particle).  

However, it overlooks the 
curious fact that the electromagnetic field produced by a particle
uniformly accelerated for all time (the retarded solution to the
distributional Maxwell equations)
does {\em not} vanish on this plane, 
and in fact is highly singular there. 
This can be seen in Boulware's expression \cite{boulware}[equation (III.11)]
for the field, which contains a $\delta$-function $\delta(x+t)$.

We do not agree with Singal's argument that it is legitimate to 
omit the plane $t = 0, x = 0$ from the integration on the grounds 
that no radiation
from the particle can reach this plane.  We believe that 
it is inconsistent to omit the delta-function
on $x = 0$ from the fields for the purpose of 
computing the field energy.  
Although no {\em light ray} from the particle can reach this plane, 
the delta-function on the plane 
in Boulware's expression (III.11) shows that
``radiation'' in the form of nonzero fields 
does in fact reach this plane---the delta-function field would not be there 
if the particle were not there! 

The delta-function 
is part of the particle's field, 
and it should not be surprising that if it is omitted in the energy
calculation, the result is less energy than expected at $t=0$. 
Intuitively, at $t=0$, all the radiated energy is concentrated in the
delta-function on the plane $x=0$.    

We cannot prove this by completing the calculation 
because the standard expression for the field
energy entails squaring this $\delta$-function, a mathematically
ill-defined operation which leads to seemingly mathematically
meaningless expressions like the $\delta(0)\delta(x+t)$
factor in Boulware's equation (IV.2) \cite{boulware} for the
component $T^{tt}$ of the energy-momentum tensor 
whose integral over space would normally give the energy in the plane.
Unfortunately, \cite{boulware} does not furnish 
a mathematically meaningful interpretation for this expression. 

\section{Discussion of related paradoxes}

It may be helpful to compare Singal's paradox with a similar
``paradox'' whose resolution is well understood. 
In this ``paradox'', two plausible methods for calculating 
a (not necessarily uniformly) accelerated particle's energy radiation,
one due to Dirac \cite{dirac} and the other to Bhabha \cite{bhabha},
give the same answer for an asymptotically free particle 
(such as one accelerated for only a finite time),
but usually different answers 
for a particle which is not asymptotically free
(in particular for a particle uniformly accelerated for all time).  

It seems generally accepted that a charged particle which undergoes 
acceleration (not necessarily uniform) {\em for only a finite time} 
does radiate in accordance with the Larmor law.  Specifically,
if a particle of charge $q$ 
with four-velocity $u^i$ and acceleration $a^i := du^i/d\tau$
is free (i.e., unaccelerated) for 
proper times $\tau \leq \tau_1$ and $\tau \geq \tau_2$, then
it radiates total energy-momentum $P^j$ given by: 

\begin{equation}
\label{bhabharad}
P^j = - 2q^2 /3 \int_{\tau_1}^{\tau_2} a^i a_i u^j \, d\tau
\q .
\end{equation}
Here vector components are with respect to an orthonormal basis
for Minkowski space with metric $g_{ij} = \mbox{diag}(1,-1,-1,-1)$,
and the energy radiation is $P^0$.

We emphasize that the hypothesis under which \re{bhabharad}
is unequivocally accepted is that the particle be free 
outside the proper time interval $\tau_1 \leq \tau \leq \tau_2$.
We shall call this the ``fundamental hypothesis''.%
\footnote{
Space limitations preclude a proper treatment 
of this important point here, but 
a general discussion can be found in Chapter 4 of \cite{parrott},
along with full definitions of the Bhabha and Dirac ``tubes'' 
to be briefly described below and the associated
radiation calculations. 
Discussion of this point as it applies specifically
to uniform acceleration can be found in the Internet archive \cite{parrott2}. 
}
The radiation expression \re{bhabharad} 
is accepted because all accepted calculational methods 
seem to lead to this conclusion under the fundamental hypothesis.  

A typical calculation might surround the particle's worldline
by some sort of three-dimensional tube and obtain $P^j$ as the 
integral of the Hodge dual of the 
energy-momentum tensor ${T^i}{_j}$ over this tube.  
For example, this is how Dirac originally obtained the
Lorentz-Dirac equation \cite{dirac}.

Expressed in three-dimensional language, 
such a calculation surrounds the particle with a two-dimensional surface
(usually a sphere of a given radius $r$ relative to some 
appropriate reference frame), 
integrates the Poynting vector over the surface to obtain the rate of
energy radiation, and then integrates the radiation rate over all time
to obtain the total radiation.   

There are a number of reasonable ways to choose the tube
or surface, but the fundamental hypothesis
guarantees that the result $P^j$ of the calculation is the same for
all accepted methods known to this author.   
This is because the covariant divergence $\partial_i T^{ij}$ vanishes.  
Two different tubes can be smoothly joined at the ends 
{\em assuming that the acceleration vanishes at the ends},
and then the vanishing divergence together with Gauss's Theorem 
implies that the calculated $P^j$ is independent of the tube.

Many of the seeming contradictions 
in the literature of the problem of
radiation from a uniformly accelerated charge 
can be traced to overlooking the ``fundamental hypothesis''.  
For example,
Bhabha \cite{bhabha} calculated the radiated energy-momentum from
an arbitrarily accelerated particle using 
a tube built from ``retarded spheres'' $S_\tau$ 
(obtained by imagining the particle emitting a flash of light at proper
time $\tau$ and letting it expand to a sphere of a given radius $r$ relative 
to the particle's rest frame at $\tau$), and the result was the 
above $\re{bhabharad}$. 
Dirac \cite{dirac} used a different tube 
generated by evolving through time $\tau$ 
two-dimensional spheres $\bar{S}_\tau$ of radius $r$
with the particle at the center, 
where the spheres are taken relative to the  
particle's rest frame at proper time $\tau$. His calculation yielded
(in the limit $r {\scriptstyle \rightarrow}  0$)
the different result 
\begin{equation}
\label{diracrad}
P^j = - 2q^2/3 \int_{\tau_1}^{\tau_2} (a^i a_i u^j  + da^j / d\tau) \, d\tau 
\q .  
\end{equation} 
The derivations of both of these results are of a high standard of rigor,
and seem universally accepted.  
They obviously agree under the fundamental hypothesis,
but ignoring this hypothesis results in a ``paradox''. 

For uniform acceleration 
(meaning that the acceleration is in a fixed spatial direction
with $a^i a_i$ constant 
and $\tau_1 := - \infty$, 
$\tau_2 := \infty$), the Bhabha and Dirac methods do {\em not} agree. 
In this case, 
the integrand of \re{diracrad} 
is well known to vanish identically 
\cite{fulton/rohrlich} \cite{parrott2},
%
%
but the energy component of \re{bhabharad} 
is strictly positive when $a \neq 0$. 
Thus for uniform acceleration for all time,
the Dirac calculation gives zero energy radiation, 
while the Bhabha calculation gives infinite energy radiation.

Obviously, both cannot be correct.  The only known resolution of  
this ``paradox'' is that the usual justifications for these  
methods (in particular, mass renormalization)
require that  $a(\tau_1) = 0 = a(\tau_2)$,
and since the methods are of comparable plausibility 
and give inconsistent results in the absence of this assumption, 
both should be disallowed.  

\section{Conclusions}

So far as we know, no one questions that a charged particle which is
uniformly accelerated for a finite time does radiate energy. 
In particular, Singal's method yields the usual Larmor radiation expression
for this situation.

However, some calculational methods, including Singal's, predict zero energy
radiation when applied to a particle uniformly accelerated for all time.
Other equally plausible methods (such as integration over a Bhabha tube)
predict the contradictory result of infinite radiation 
in accordance with the Larmor law. 

We suggest that uniform acceleration for all time
should be recognized as too mathematically singular to be treated reliably with 
the sort of mathematical manipulations customary in this field. 
One indication of this is the delta-function in the field at a point
not causally connected to the particle and the seemingly mathematically 
meaningless term in the energy-momentum tensor corresponding to its square. 

The situation might be compared to ambiguities in calculation of 
limits which perplexed 19'th century mathematicians and 
spurred the development of rigorous foundations for calculus.
For example, if the series $\sum_{n=1}^\infty (-1)^n $ is summed as
$\ [-1 + 1] + [-1 + 1] + \cdots\ $, 
the result is 0, whereas
summing it as 
$\ -1 +[1 + (-1)] + [1 + (-1)] + \cdots\ $  
gives $-1$.  
If one attempts to apply algebraic rules valid for finite sums
to infinite sums 
without any proof of their validity, 
one can expect to obtain such seemingly paradoxical results.  
There is much evidence that a charge uniformly accelerated forever 
is similarly a mathematically singular situation 
in which algebraic manipulations conventional in mathematical physics 
can be expected to lead to contradictory results.  
\vspace{1ex} 

\noindent 
{\bf Acknowledgement:} 
I thank Ashok Singal for reading an earlier version of this paper
and spotting a calculational error.  

\end{document}